\documentstyle[aps,prb,multicol,epsf,graphicx]{revtex}

\begin{document}

\title{A real-space description of the glass transition based on
 heterogeneities and entropy barriers}

\author{Andrea Crisanti$^{1}$ and 
 Felix Ritort$^{2}$} \address{$^{1}$Dipartimento di Fisica, Universit\`a di Roma ``La Sapienza'',
         P.le Aldo Moro 2, I-00185 Roma, Italy \\
         Istituto Nazionale Fisica della Materia, Unit\`a di Roma}
 
\address{$^{2}$Departament de F\'{\i}sica Fonamental,
 Facultat de F\'{\i}sica, Universitat de Barcelona\\ Diagonal 647,
 08028 Barcelona (Spain)\\ E-Mail: crisanti@phys.uniroma1.it,ritort@ffn.ub.es}


\maketitle
\begin{abstract}
An alternative scenario for the glass transition based on the
cooperative nature of nucleation processes and the role of entropic
effects is presented. The new ingredient is to relate the dissipation
during the relaxation process to the release of strain energy driven
by the nucleation of progressively larger cooperative spatial
regions. Using an equiprobability hypothesis for the transition between
different metastable configurations, we obtain a relation between the
free energy dissipation rate and the size of the largest cooperative regions. This leads to a new
phenomenological relation between the largest relaxation time in the
supercooled liquid phase and the effective temperature. This differs
from the classical Adam-Gibbs relation in that predicts no divergence
of the primary relaxation time at the Kauzmann temperature but a
crossover from fragile to strong behavior.
\end{abstract}
\begin{multicols}{2}
\narrowtext

{\it Introduction.} After many decades of efforts, a complete and
unifying description of the glass transition problem is still not
available. The standard approaches to the glass transition have been
largely based on hydrodynamic (such as MCT \cite{MCT}) or
thermodynamic entropic theories like that proposed by Adam, Gibbs and
Di Marzio (hereafter referred to as AGM) nearly 50 years ago
\cite{AGM}. Surprisingly, still nowadays the ideal AGM theory remains
not accepted nor disproved. A salient outcome of the AGM theory is the
prediction of the existence of a second-order phase transition driven
by the collapse of the configurational entropy (also called
complexity). It is known that an unambiguous definition of the
complexity is possible only in the framework of mean-field theories
where phase space splits into ergodic components of infinite
lifetime. Recent approaches to the glass transition problem from the
perspective of disordered systems have, to a large extent, validated
this mean-field scenario \cite{KTW}. Nevertheless, a complete
understanding of the glass transition must go beyond mean field by
including nucleation processes into any valuable theory. This poses
the question about how the mean-field picture for the glass transition
is modified in the presence of real-space effects.  In this paper we
propose an alternative scenario for the glass transition where spatial
effects such as heterogeneities play a crucial role in the theoretical
description of the glassy state being also the necessary ingredient to
validate and reinforce some aspects of entropic mean-field
approaches. We present a phenomenological real-space description of
nucleation processes of cooperative nature in the spirit of the AGM
theory. This should complement other approaches such as mode-coupling
theory \cite{MCT}. The main outcome of our analysis is that the
standard relation between the relaxation time and the complexity
originally proposed in the AGM theory is modified and the primary
relaxation time does not depend directly on the complexity but on the
so called {\em effective temperature} (to be defined later).  This
predicts a crossover from fragile to strong behavior and a saturation
(and not a divergence) of the effective barrier or primary relaxation
time at the Kauzmann temperature.

{\it A phenomenological description of heterogeneities in glasses.}  One
of the most intriguing features in glasses is the existence of
heterogeneous structures \cite{HETERO}. Experimentally these manifest as
some set of atoms which have a dynamics manifestly slow when compared to
the rest. Although experiments or numerical simulations in this field
are very recent, heterogeneities are a direct manifestation of the
cooperative nature of nucleation processes \cite{IS}. The basic idea in
our approach is to assume that nucleation processes take place
everywhere inside the glass when some structures of size $s$ are built
up by a cooperative mechanism. To built these structures the $s$ atoms of
the cooperative region must coherently move to occupy certain positions
which enable that region to release some strain energy. Each of these
moves constitutes an elementary activated process, for instance, the
exchange between two neighboring particles. Therefore, the
heterogeneities observed in the experiments are transient frozen
structures which eventually nucleate in time scales larger than the
observational time. When a droplet of liquid nucleates it changes to a
new locally disordered structure. The local structure of the glass is
always that of a liquid and there is no coarsening of a given pattern
whatsoever \cite{KT}. In our scenario the time to activate a region of size $s$ is,

\begin{equation}
\frac{\tau(s)}{\tau_0}\propto \bigl(\frac{\tau^*}{\tau_0}\bigr)^s=\exp\bigl(\frac{Bs}{T})
\label{eqxi}
\end{equation}

\noindent
where $\tau^*=\tau_0\exp(B/T)$ is the activated time to activate one
atom, $\tau_0$ being a microscopic time and $B$ the corresponding
energy barrier. Let us introduce the quantity $n_s(t)$ as the number
of cooperative regions (we will refer to them as domains) of size $s$
at time $t$. Experimentally it is well known that time correlations in
the glass state are stretched but decay faster than any power
law. According to (\ref{eqxi}) this means that the distribution $n_s$
must abruptly fall down beyond a cuttof size $s^*$ in such a way that
long time nucleations occur with a negligible
probability. Consequently, $n_s(t)$ for $s>s^*$ must be nearly zero. This
approach is similar in spirit to the mosaic theory developed by Xia
and Wolynes \cite{WX} who also considered the existence of a cutoff
size $s^*$. The existence of this cutoff is tightly related to the
cooperative character of the dynamics itself eq.(\ref{eqxi}) and can
be illustrated within a simple domain aggregation model. We note that
the model we present here is oversimplified, our aim being only to
stress the relation between cooperativeness and the presence of a
cutoff size. Let us imagine a liquid that is quenched to a temperature
$T_f$ where equilibration cannot be achieved within the time scale of
the experiment. Let us think about the molecules inside the glass as
grouped into non-overlapping domains of different sizes $s$. After
nucleating, domains of a given size $s$ destabilize breaking into
smaller domains. In the simplest scenario the
aggregation and breaking of domains occurs inside a bath of
particles. We assume that, after nucleation of a domain of size $s$,
domains can gain or loose one particle with respective
probabilities $g_s,l_s$ with $g_s+l_s=p_s$. $p_s$ is the probability
that a nucleation occurs which we take proportional to $1/\tau(s)$
where $\tau(s)$ is given by eq.(\ref{eqxi}) with $B=\tau_0=1$ and
$T=T_f$. For sake of simplicity we take $g_s=gp_s,l_s=lp_s$ with
$g+l=1$. Consequently, the balance equations involve the following
``chemical reactions'': ${\cal D}_s \rightarrow {\cal D}_{s-1}+ p$
with rate $l_s$ and ${\cal D}_s+p \rightarrow {\cal D}_{s+1}$ with
rate $g_s$, where ${\cal D}_s$ denotes a domain of size $s$ and $p$ a
single particle of the bath. The balance equations read ($s\ge 2$),

\begin{equation}
\frac{d n_s(t)}{d t}=
l_{s+1}n_{s+1}(t)+g_{s-1}n_{s-1}(t)-p_sn_s(t)~~~.
\label{eq_balance}
\end{equation}

This set of equations must be supplemented with the dynamical equation
for the ``bath'' of particles which is deduced from mass conservation
$\sum_{s=1}^{\infty}sn_s(t)={\rm const}$. In this simple model there are
three parameters $g,l$ and $\beta=1/T_f$ entering the equations. By
appropriately rescaling the time only two of them ($\beta,g/l$) are free
parameters. Although all possible values of $g,l$ are possible the
interesting regime is obtained for $g/l$ small. Physically this means
that, during the nucleation, domains have more probability to loose
particles than to capture them. This is a very reasonable assumption:
just before the nucleation takes place the domain is in an unstable
configuration and loosing a particle seems a more probable event. We
numerically solved (\ref{eq_balance}) by numerically integrating them
using a second order Euler algorithm.  In the left panel of figure 1 we
show the time evolution for $n_s(t)$. At any time it displays a well
defined time dependent cutoff value $s^*(t)$ above which the $n_s(t)$
drops to zero very fast.  From the knowledge of $n_s(t)$, assuming
independent exponential relaxations for the different domains, we can
also obtain the two-time correlation function,

\begin{equation}
C(t,t+t')=\sum_{s\ge 1} s n_s(t')\exp(-t/\tau(s))
\label{corr}
\end{equation} where $t$ denotes the time after the quenching and $\tau(s)$ is
given by expression (\ref{eqxi}). The correlations can be excellently
fitted by a stretched exponential law with a $t$-dependent stretching
exponent $\beta_s$, $C_{t}(t')=\exp(-(t'/\tau(t))^{\beta_s(t)}$. In the
right panel of figure 1 we show the corresponding correlation functions
as well as the best fits.  Note that the average relaxation time
$\tau_{av}=\int_0^{\infty}dt'C_{t}(t')$ does not necessarily scale like
$t$. A careful examination of the time dependence of the distribution
$n_s(t)$, reveals that it scales like $n_s(t)=(1/s^*)\hat{n}(s/s^*)$
with $t=\exp(\beta s^*)$, $s^*(t)$ being the time
dependent cutoff size.  Consequently, relaxation to equilibrium is
driven by the growth of the largest domain of size $s^*(t)$. When a
glass is relaxing to equilibrium, the average release of energy to the
thermal bath is done by nucleation of the largest domains of size $s^*$,
meaning that the advance motion of the {\it front} of $n_s(t)$ located
at $s=s^*$ is the leading source of energy dissipation. Nucleation
processes involving regions of size smaller than $s^*$ always occur but
do not yield a net energy current flow to the thermal bath. Metastable
equilibrium is reached when the cutoff size $s^*$ saturates to a finite
value and the net energy flow between the glass and the bath vanishes.

\begin{figure}[tbp]
\begin{center}
\rotatebox{270}{
\includegraphics*[width=6cm,height=9cm]{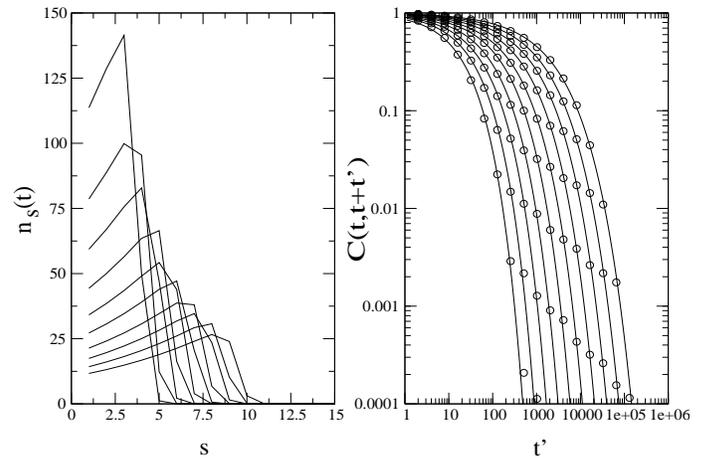}}

\vskip 0.05in
\caption{$n_s(t)$ and $C(t,t+t')$ for different times
$t=10^{14}-10^{33}$ (in adimensional units) for the numerical
solution of model (\ref{eq_balance}) with $l=8,g=1$ and
$\beta=2.2$. The parameter $\tau(t)$ and the stretching exponent are very
well fitted by $\tau(t)=2.2t^{0.35}$,
$\beta_s(t)=0.34+0.45t^{-0.06}$.
\label{fig1}}
\end{center}
\end{figure}      

Let us now focus our attention in the energy dissipation rate in this
relaxation process. This is given by typical free energy variation of
the relaxing domains of size $s^*(t)$ multiplied by
their number $N_{s^*}(t)=V n_{s^*}(t)=(V/s^*)\hat{n}(1)$, divided by
their typical nucleation time $t=\exp(B\beta s^*)$,

\begin{equation}
\frac{1}{V}\frac{\partial F}{\partial t}\sim \frac{\Delta
F^* N_{s^*}(t)}{Vt}\sim \frac{\Delta F^*}{s^* t}~~~~.
\label{dissip1}
\end{equation}

Now, the crucial difference between nucleation processes in glasses
and other systems is the fact that nucleation in a glass
occurs between metastable liquid drops \cite{AGM}. The typical release
of energy $\Delta F^*$ released after nucleating a metastable liquid
drop of size $s^*$ into another drop does not scale with its surface
(like in coarsening systems) nor its volume (like in standard
liquid-solid nucleations) but is finite and independent of $s^*$
yielding,

\begin{equation}
\frac{1}{V}\frac{\partial F}{\partial t}\sim-\frac{\Delta_F}{s^* t}~~~~.
\label{dissip1_b}
\end{equation}

\noindent where $\Delta_F$ is the average free energy change
\cite{AGM}.  The growth of $s^*$ as a function of time is $s^*(t)\sim
T\log(t)$ and stops when $n_s(t=t_{\rm eq})$ reaches the stationary
distribution. In principle the value of the equilibration time $t_{\rm
eq}$ depends on the particular model under consideration. From a
microscopic theory this time can be deduced only from the hydrodynamic
equations. We see below how we can circumvent the hydrodynamic
description by relating the relaxation time of the supercooled liquid
to the complexity.

{\it The fluctuation formula.} One of the main problems in the
theory of glasses is to define the complexity $S_c$. In the
original AGM theory, $S_c$ was defined as the configurational part of
the entropy obtained by disregarding the kinetic part of the energy.
Here we consider a different dynamical definition adapted from concepts
developed in spin-glass theory \cite{THEO,KB}. The key concept in our
definition is the concept of conformation or basin: a basin includes
those configurations which do not release an extensive (proportional
to $s$) strain energy to the thermal bath after nucleating a region of
size $s$. Let us consider a relaxing liquid quenched to
$T_f$. We define $S_c(t,F)$ as the logarithm of the total number of
basins\cite{FOOTNOTE} with free energy $F$ which can release strain energy to
the thermal bath after nucleating regions of size $s^*$ at time
$t$. In other words, because $s^*$ is the maximum size of nucleating
regions at time $t$, $S_c(t,F)$ counts the number of still {\it not
visited} conformations at time $t$. In this scenario regions of size
$s$ which have already nucleated once (i.e. $s<s^*$) have already
released their strain energy while regions which have not yet nucleated
($s>s^*$) still contain some strain energy. Nucleations which do not
lead to a new release of strain energy do not yield new conformations.

Each of these conformations at
time $t$ may contain different possible configurations ${\cal C}$
which do not contribute to the complexity but contribute to the free
energy of the conformation ${\cal B}$, $F_{\cal
B}(t,T_f)=-T_f\log(\sum_{{\cal C}\in {\cal B}} \exp(-E({\cal
C})/T_f)$ where time dependent expectation values of
a given observable $A_{\cal B}(t,T_f)$ are defined as follows: $A_{\cal
B}(t,T_f)=\sum_{{\cal C}\in{\cal B}}A({\cal
C})\exp\Bigl(-\frac{(E({\cal C})-F_{\cal B}(t,T_f))}{T_f}\Bigr)$.

The liquid character of the glass phase implies that basins with
identical free energy $F_{\cal B}(t,T_f)$ have reproducible physical
properties (i.e., independent of ${\cal B}$).  Note that the free energy
and the expectation value of any observable $A$ evaluated at a given
conformation ${\cal B}$ depend on both $T_f$ and the quenching time
$t$. The dependence on $t$ directly appears through the set of
configurations ${\cal C}$ contained in ${\cal B}$. In particular, for
large enough times, basins with very high free energy only contain
configurations which have been explored several times, hence they do not
contribute to the complexity. Contrarily, conformations with very low
free energy contain configurations which have still not been reached,
hence they contribute to the complexity. This leads to a dynamical
complexity $S_c(t,F)$ which has a t-dependent cutoff value $F^*$ such
that the number of conformations or basins with $F_{\cal B}(t,T_f)>F^*$
vanishes.  Moreover, the complexity is a monotonous increasing function
of the free energy since the number of possible conformations which
decrease the energy $N_{s^*}(t)=(V/s^*)\hat{n}(1)$ is larger when
nucleating regions are smaller. In the asymptotic large $t$ limit it
takes the form,

\begin{equation}
S_c(t,F<F^*)=\hat{S_c}(F,T_f)~;~S_c(t,F>F^*)\to -\infty
\label{Sc}\end{equation}

 Now we can introduce the fluctuation formula and see how
the present scenario substantially differs from the classical one by
AGM. At time $t$ after the quenching, nucleation processes inside the
glass occur between cooperative regions with a disordered local
structure and no characteristic pattern grows with time. This amounts to
say that nucleations must be entropically driven meaning that some sort
of equiprobability hypothesis for visiting conformations holds. While
this is an assumption, its validity must be founded on principles
similar to those which justify the equiprobability hypothesis in
Boltzmann-Gibbs equilibrium theory. Therefore, the
probability to jump from the free energy level $F^*$ to another
conformation with free energy $F$ is always proportional to the number
of configurations with final free energy $F$,

\begin{equation}
{\cal W}_{F^*\to F}\propto \frac{\Omega(F)}{\Omega(F^*)}\propto
\exp(\hat{S_c}(F,T_f)-\hat{S_c}(F^*,T_f))~~~.
\label{W}
\end{equation}

Substituting (\ref{Sc}) in (\ref{W}) and denoting $\delta F=F-F^*$ 
we obtain for the transition probability,

\begin{equation}
{\cal W}_{\delta F}\propto\exp\Bigl(\beta_{\rm eff}(F^*)\delta
F\Bigr)\theta(-\delta F)
\label{FF}
\end{equation}

where we have defined the effective temperature $T_{\rm
eff}(F^*)=1/\beta_{\rm eff}(F^*)$ with $\beta_{\rm
eff}(F^*)=\bigl(\frac{\partial \hat{S_c}(F,T_f)}{\partial
F}\bigr)_{F=F^*}$ being the density of complexity with free energy $F^*$.
This formula establishes the probability of free energy jumps in the
aging state. Fluctuations to conformations or basins which increase
the free energy are forbidden, simply because they have already
nucleated in the past. While fluctuations to very low free energy
conformations are entropically suppressed due to the monotonous
increasing property of $\hat{S_c}(F,T_f)$.  In the off-equilibrium state
the average rate variation of the free energy at time $t$ is then given
by,

\begin{equation} \frac{1}{V}\frac{\partial F}{\partial t}\sim
-\frac{\int_{-\infty}^{\infty}x{\cal
W}_xdx}{t\int_{-\infty}^{\infty}{\cal W}_xdx}= -\frac{1}{\beta_{\rm
eff}(F^*) t}~~~~.
\label{dissip1_c}
\end{equation}

We can now establish the connection between equation (\ref{dissip1_b})
and the present one. Consistency requires that $s^*=\Delta_F\beta_{\rm
eff}(F^*)$, i.e. the size of the largest nucleating regions $s^*$ is
directly proportional to the inverse effective
temperature $\beta_{\rm eff}(F^*)$ evaluated at the time-dependent free
energy $F^*$.  After equilibrating at time $t_{\rm eq}$, $F^*$ converges
to $F_{\rm eq}$ and $s^*$ has saturated to a value $s^*(t_{\rm eq})$ which
determines the new phenomenological relation,

\begin{equation}
\tau_{\alpha}=\tau_0\exp\Bigl (\frac{Bs^*(t_{\rm
eq})}{T_f}\Bigr)=\tau_0\exp\Bigl(\frac{B\Delta_F\beta_{\rm eff}(F_{\rm eq})}{T_f}\Bigr).
\label{CR}
\end{equation}

We can now compare our prediction with the standard AGM scenario
\cite{AGM}. In that scenario the number of different conformations at
time $t$, $\Omega_{t}$ corresponds to all possible combinations
obtained from the two possibilities (nucleated and not
yet nucleated) for the largest nucleating regions $n_{s^*}(t)$. This
number is given by $\Omega_{t}=2^{\frac{V}{s^*}}$ yielding for the
complexity, $S_c(t)=\log(\Omega_{t})=V(\log(2)/s^*)$ which relates the
size of the cooperative region to the complexity. Using eq.(\ref{eqxi})
and taking $t=t_{\rm eq}$ this yields the famous Adams-Gibbs relation
$\tau_{\alpha}=\exp(B\log(2)V/T_fS_c)$. The strong assumption contained
in the AGM relation is to suppose that, in the off-equilibrium regime
during the relaxation, the glass explores nucleated and not nucleated
conformations with the same probability. But this cannot be true if free
energy jumps are biased towards lower free energy
conformations. Actually, according to AGM, the size of the cooperative
region increases with time and diverges at the Kauzmann temperature
$T_K$ while in our theory it saturates to a finite value.

One more consequence of the fluctuation formula (\ref{FF}) concerns the
fluctuation-dissipation ratio (FDR) and its one-step character
\cite{CK,SIM,ROM}. After quenching to $T_f$ a possible way to quantify
violation of FDR is to measure the average value of any observable $A$
after a perturbation field $h_A$ conjugated to the observable $A$ is
applied to the system. Due to the disordered structure of the
cooperative regions, if the perturbation equally weights all
conformations, the entropically driven assumption implies that states
with free energy $F$ are sampled with a probability proportional
to their number. In the presence of a field $h_A$ the complexity must be
a function of three variables, $\hat{S_c}(F,T_f,A)$. We can simply
obtain the average change in the expectation value $<A(t)>$ after
switching on the perturbation field $h_A$ at time $t_w$. In the linear
response regime \cite{FV} the Onsager regression principle implies for
the transition probabilities ${\cal W}_{\delta F,\delta A=A-A_0}={\cal
W}_{\delta F}\exp(\hat{S_c}(F,T_f,A)-\hat{S_c}(F,T_f,A_0))={\cal
W}_{\delta F}\exp(\frac{\partial \hat{S_c}(F,A)}{\partial
A})_{A_0}$. Using the relation $(\frac{\partial \hat{S_c}(F,A)}{\partial
A})_{A_0}=-\beta_{\rm eff}(F)h_A$ and expanding for $h_A$ small we
finally get the famous violation FDT expression, $\frac{\partial
<A>(t)}{\partial h_A(t_w)}=\beta_{\rm
eff}(F^*)(<A(t)(A(t)-A(t_w))>_{h_A=0}$. Note that the description of the
violation of FDT in terms of a single time scale $t$ is consequence of
the asymmetric shape (i.e. ${\cal W}_{\delta F>0}=0$) of (\ref{FF}).

{\it Main implications of the present scenario.} It can be proven
\cite{MARC} that $\beta_{\rm eff}(F)\le 1/T_K$ where $\hat{S_c}(T_K)=0$
at $T_K$. This implies an asymptotic crossover
for all fragile liquids to strong behavior. A strong glass is a fragile
one which has exhausted all its complexity and the effective barrier has
saturated to its maximum value $\beta_{\rm eff}=1/T_K$. Instead fragile
glasses have high excess complexity and still big variation of
$\beta_{\rm eff}$ along the supercooled line. The non-trivial
temperature dependence of the activation barrier in (\ref{CR}) through
the non-universal quantity $\beta_{\rm eff}$ explains why it is so
difficult to find a unique empirical law that properly describes the
viscosity anomaly of all glasses. The prediction that $s^*$ saturates to
a finite value $s^*=\Delta_F/T_K$ at $T_K$ is not easy to check
experimentally due to the difficulty to cover one order of magnitude in
the effective barrier. Yet some experimental results suggest a crossover
from fragile to strong behavior\cite{CAVAGNA,CROSSOVER}. The present
scenario could be also checked doing numerical aging
experiments. According to (\ref{CR}) the $t$ dependent
effective temperature $\beta_{\rm eff}(t)$ (measured through
FDT-violations or the formula \ref{FF}) should be given by $\beta_{\rm
eff}(t)\to(T_f/B\Delta_F)\log(t/\tau_0)$ if
$\tau_{\alpha}(t)\simeq t$ and both $\tau_0$ and $B\Delta_F$ are
nearly $t$ and $T_f$ independent. This relation should hold for all
$T_f$ and $t$ predicting a value for the activation barrier and the
cooperative size \cite{ROM}. The experimental confirmation of a
saturation of the heterogeneity sizes for not too fragile glasses and
the crossover from fragile to strong behavior is probably not out of
reach and would be a check of the validity of the present theory.

{\bf Acknowledgments}. We wish to thank suggestions by C. Cabrillo, G. Parisi,
F. Sciortino and G. Tarjus. F.R. is
supported by the MEC in Spain, project
PB97-0971.
\hspace{-2cm}

\end{multicols}

\begin{thebibliography}{99}
\bibitem{FOOTNOTE} Actually, to avoid confussion it would be better to call
$S_c$ conformational or basin entropy rather than configurational entropy.
\end{thebibliography}
\end{document}